\documentclass{aa}
\usepackage{graphicx,psfig,amssymb,natbib}

\def\cunits{cm$^{-2}$}

\def\lunits{erg~s$^{-1}$}
\def\mbh{$M_{\bullet}$}
\def\msun{$M_{\odot}$}
\def\msuny{\msun yr$^{-1}$}

\def\cs{$\chi^2$}
\def\csn{$\chi^2_\nu$}
\def\fx{$f_X$}

\def\lbol{$L_{\rm bol}$}
\def\ledd{$L_{\rm Edd}$}

\def\logxr1{$\log (f_X/f_R) \le -1$}

\def\lx{$L_X$}

\def\lxol2{$\log (f_X/f_O) < -2$}

\def\nh{$N_{\rm H}$} 
 
\def\nhgal{$N_{\rm H}^{\rm gal}$} 
\def\nhint{$N_{\rm H}^{\rm int}$} 
\def\ran{$-2 < \log (f_X/f_O) < -1$}
\def\ran1{$-1 < \log (f_X/f_O) < +1$}

\def\x{X-ray}

\def\feka{Fe K$\alpha$}

\def\ha{H$\alpha$}
\def\hb{H$\beta$}

\def\hone{H{\sc \,i}}
\def\htwo{H{\sc \,ii}}

\def\ntwof{[N{\sc \,ii}]}

\def\othreef{[O{\sc \,iii}]}

\def\oonef{[O{\sc \,i}]}

\def\stwof{[S{\sc \,ii}]}

\def\chandra{{\it Chandra}}

\def\hst{{\it HST}}

\def\rosat{{\it ROSAT}}
\def\2df{{\it 2dfGRS}}
\def\xmm{{\it XMM-Newton}}

\def\fr{Fig.~\ref}

\def\tr{Table~\ref}

\def\exi{\begin{equation}}
\def\exo{\end{equation}}

\newcommand{\aer}[3]{$#1^{#2}_{#3}$} 
\newcommand{\ten}[2]{$#1\times 10^{#2}$} 

\def\spose#1{\hbox to 0pt{#1\hss}} 
\def\approxlt{\mathrel{\spose{\lower 3pt\hbox{$\sim$}}
        \raise 2.0pt\hbox{$<$}}}
\def\approxgt{\mathrel{\spose{\lower 3pt\hbox{$\sim$}}
        \raise 2.0pt\hbox{$>$}}}

\begin{document}

\title{Searching for hidden AGN in nearby star-forming galaxies with \chandra.} 

\author{P.~Tzanavaris \and  I.~Georgantopoulos}
\offprints{P.~Tzanavaris; email: pana@astro.noa.gr}

\institute{Institute of Astronomy \& Astrophysics, \ National
Observatory of Athens, I.~Metaxa \& V.~Pavlou, Penteli 15236, Greece}
\date{Received... ; accepted...}

\abstract
{}
{
We searched for X-ray signatures of AGN in a sample of star-forming, 
relatively early-type, nearby
spiral galaxies, from the optical spectroscopic sample
of Ho et al. The tight correlations between the masses of
central supermassive
black holes and bulge mass or velocity dispersion suggest that
such galaxies are likely to host black holes in their centres.
The aim is to look for X-ray signs of activity in 
these gas-rich environments. 
}
{
We identified \chandra\ ACIS-S images of the X-ray counterparts
of 9 sources from the optical sample. We isolated 10 individual X-ray sources
closest to the optical position and calculated their X-ray luminosity, \lx, 
and \x\ colours. Using \ha\ luminosities, we also calculated
star formation rates.
For four sources with a high number of net counts in the X-ray band, we
extracted X-ray spectra to which we fitted standard spectral models.
We modelled \feka\ emission by Gaussians, and assessed the
significance of adding such a component to a fit by
means of a calibration of the standard $F$-test. 
For the rest of the sources, we estimated values for
the intrinsic hydrogen column density, \nhint, and the power-law photon index, $\Gamma$, which can
reproduce the observed soft and hard \x\ colours.
}
{
All spectral fits include a power law, a diffuse hot gas, 
a Galactic absorption and
an intrinsic absorption component. For the nuclear sources of NGC 2782 and
NGC 3310, a \feka\ emission-line component is included with
high significance. At the same time the power-law
component is flat (index $\Gamma \sim 0$ and 1.3).
Both sources have high star-formation rates, with the rate
for NGC 2782 being the highest in our sample.
}
{
The detection of \feka\ emission coming from a central,
isolated source in NGC 2782 and NGC 3310, points towards
a hidden AGN in two out of nine relatively early-type, star-forming
spirals observed both with Palomar and \chandra. Larger samples need to
be explored, but this evidence suggests that the presence of
an AGN is likely in massive, disturbed starbursts. The
contribution of X-ray emission from these AGN is small: For the
central source, the 
bolometric luminosity is a fraction of a \ten{\rm few}{-5} of the
Eddington luminosity. The X-ray flux of the central source is between
10 and 20\%\ of the total galactic flux in the $0.2-10$~keV band.
}

\keywords{X-rays:galaxies--
Galaxies:starburst-- Galaxies:active-- Galaxies:individual (NGC 2782)--
Galaxies:individual (NGC 3310)} 

\titlerunning{Hidden AGN}
\authorrunning{P.~Tzanavaris et al.}  

\maketitle

\section{Introduction}
The main energy source of X-ray emission from galaxies with an active
galactic nucleus (AGN) is gravitational potential energy, released as
gas accretes onto a central super-massive black hole \citep[SMBH,
][]{1986seag.proc..447R,1969Natur.223..690L,1964ApJ...140..796S}. 
This process leaves an imprint on X-ray 
spectra in the form of a power-law, perhaps due to inverse Compton
up-scattering in a hot corona \citep[][and references therein]{2006Rey},
although the process is not fully understood.

A number of studies suggest that {\it all} galaxies with bulges should
harbour central black holes with masses that scale with bulge mass, as
inferred from bulge luminosity or stellar velocity dispersion
\citep[e.g.,][]{1998AJ....115.2285M,2000ApJ...539L..13G}.  However,
only a small fraction of bulge galaxies actually show evidence for AGN
activity. For instance, in about half of the high signal-to-noise
ratio optical spectra taken by \citet[][henceforth
HFS]{1997ApJS..112..315H}, which include several early-type spirals,
there is no indication of AGN activity.  The straightforward
conclusion that all central black holes are dormant is puzzling, given
that these are very gas-rich environments. It is possible that,
although low-level AGN activity is actually taking place in these
systems, its optical spectral signatures are drowned by those of
circum-nuclear star formation. 
X-ray observations have proved to be an excellent tool for
detecting the presence of a hidden AGN.
In particular,
nuclear Fe line emission in AGN is a unique spectral feature produced
via X-ray fluorescence by matter surrounding a central black hole. Fe
K emission has indeed been detected in several LLAGN and
low-ionisation emission line regions
\citep[LINERs,][]{2002ApJS..139....1T,2000ApJ...535L..79T}, clinching
the evidence for a central engine.

In the last few years, using \chandra,
there have been a number of
studies of nuclear sources in {\it early-type} galaxies. In the
largest such study, \citet{2005ApJ...624..155P} examined the presence
of nuclear sources in a sample of 50 early-type (E-Sa) nearby galaxies
(most of which were from the HFS sample) and detected nuclear sources
in 39 of them, with X-ray luminosities $L_X > {\rm few}\times
10^{38}$~\lunits.  On the other hand, there are
very few {\it star-forming} systems, such as NGC
253 \citep{2002ApJ...576L..19W} and NGC 3690 \citep{2004ApJ...600..634B}, 
which may be hosting a nuclear source suggestive of AGN emission.
Similarly, 
most {\it starburst} interacting galaxies, such
as Arp 220 and the Antenn{\ae}, also turned out to be dominated
by stellar X-ray emission, without any evidence for an AGN
\citep{2001ApJ...554.1035F}. Thus, until recently, few galaxies,
optically classified as star-forming, had been found to harbour low
luminosity AGN (LLAGN), revealed at X-ray wavelengths.
However, most systems observed so far
either do not have a prominent bulge (e.g. M 82) or may be in the
process of building one, e.g.  via mergers. It is then plausible that
the central black hole may not be massive enough to emit copiously at
X-ray energies. This could then provide an explanation why, overall,
there exist no results on relatively early-type, star-forming
galaxies. For example, out of 50 galaxies in the
\citet{2005ApJ...624..155P} sample only one, NGC 660, is classified as
star-forming.

The physical and/or observational basis underlying the difficulty in
identifying two {\it clearly} distinct classes of galaxies is still
being intensively explored.  Examples of mixed AGN/non-AGN sub-classes
from the literature, which may well be related to each other, include
so-called X-ray bright optically inactive galaxies (XBONGs) and \lq
composite\rq\ galaxies.  XBONGs \citep{2005MNRAS.358..131G} have
optical spectra consistent with \lq passive\rq\ galaxies, showing no
sign of AGN activity. However, their high \lx\ values are hard to
explain without invoking an AGN
\citep{2001AJ....121..662B,2001ApJ...554..742H}.  Similarly, \lq
composite\rq\ galaxies have high \lx\ but optical spectra leading to
borderline classifications between star-forming and Seyfert types
\citep{2005ApJ...631..707P,1996ApJ...461..127M}.  This is partly
similar to cases of known AGN hosts with a significant starburst
contribution to the overall energy budget.
\citet{2001ApJ...546..845G} find that this is significant in a sample
of 20 bright nearby type 2 Seyferts. Extending this work,
these authors suggest that nuclear starbursts may well be a
general part of the Seyfert phenomenon.  Such starbursts may obscure
low-luminosity AGN (LLAGN, \lx~$\la 10^{42}$\lunits), causing a
hardening of observed X-ray spectra \citep{1998MNRAS.297L..11F}.

In the present paper we are using X-ray diagnostics to detect SMBHs
hiding in relatively early-type galaxies, optically classified as
star-forming. These are very gas-rich systems, with abundant fuel
supply to feed a central SMBH, the presence of which has not been
detected to-date.  Furthermore, estimating the level of contamination
of galactic X-ray flux by a central AGN is vital for understanding
galaxy evolution and determining the galaxy X-ray luminosity function.


Using the superb resolution of \chandra\ we
are able to isolate individual, central X-ray sources for which we extract
X-ray spectra. We then fit AGN models to these spectra and look
for Fe K emission. 




\section{The HFS optical spectroscopic sample}
HFS have carried out an extensive optical spectroscopic survey of the
nuclear regions of nearby galaxies. Their main goal was to look for,
and investigate, low luminosity AGN. Using the Palomar Observatory's
5--m Hale telescope, they obtained high-quality, long-slit, optical
spectra of 4\AA\ (2.5\AA) resolution in the spectral region
$\sim4230-5110$~\AA\ ($\sim 6210-6860$~\AA) for 486 galaxies in the
northern sky at a chosen magnitude limit $B_T=12.5$~mag. By performing
stellar template subtraction, they corrected the spectra for
contamination by absorption lines from the stellar component.  Using
the emission-line intensity ratios \othreef\ $\lambda$5007/\hb,
\oonef\ $\lambda$6300/\ha, \ntwof\ $\lambda$6583/\ha\ and \stwof\
$\lambda\lambda$6716, 6731/\ha, they identified 206 \htwo\ nuclei, 52
Seyfert nuclei, 94 LINERs and 65 transition objects.

We identified all galaxies in the HFS sample which are
classified both as
\begin{enumerate}
\item star-forming (\htwo\ nuclei), and
\item morphologically relatively early, ranging in Hubble type between
Sa and Sc, suggesting a massive bulge and, correspondingly, a
supermassive black hole.
\end{enumerate}

\section{X-ray data}
We searched the \chandra\ archive for X-ray fields observed with the
Advanced CCD Imaging Spectrometer, ACIS-S or ACIS-I, covering the
optical positions of these galaxies.  We identified nine sources,
whose positions fall within ACIS-S fields.  Details of the X-ray
observations are given in \tr{tab:X}.

We used type 2 event files (including only grades 0, 2, 3, 4, and 6)
provided by the standard pipeline processing, after discarding periods
of high background. Further, we used CIAO, version 3.2, to carry out
astrometric corrections on these data, following the standard
procedure for \chandra\ data
\footnote{\tt http://cxc.harvard.edu/ciao/threads/arcsec\_correct\linebreak 
ion/index.html}.

\section{Looking for AGN signatures}
In the X-ray band, telltale signs of AGN activity in galaxies include,
generally speaking, bright nuclear sources with high X-ray
luminosities, $\log L_X \ga 42$ (\lunits)\footnote{This is obviously not the
case for LLAGN.}, as well as the presence of the fluorescent 6.4~keV
\feka\ line.  Significant neutral hydrogen column densities,
log~\nh~$\ga 22$ (\cunits), 
are indicative of strong intrinsic absorption, often
associated with some AGN.  Our aim was to identify reliable optical
and X-ray counterparts for the nine sources mentioned and then 
search for hidden AGN by applying these criteria to X-ray
counterparts. In cases were X-ray spectra
could be extracted, we fitted spectral models to look for
\feka\ emission and obtain \nh\ estimates (\tr{tab:spec}). For the rest
of the cases, we used
the software PIMMS \citep[v3.8a,][]{1993Legac...3...21M}
to see which \nh-$\Gamma$ combination could reproduce
the observed X-ray colours, given the count rate, redshift and
galactic column density for each source.

\subsection{X-ray and optical counterparts}\label{xopt}
\citet{1999ApJS..125..409C} have determined accurate nuclear positions for 
all
confirmed galaxies in the Uppsala General Catalogue (UGC).  For each
object, they present a conservative positional uncertainty which
reflects astrometric errors, as well as errors due to galaxy size and
morphology. Positions from this work (and 1-$\sigma$ uncertainties)
are shown by the cross (and the length of its arms) in
\fr{fig:xo}. As an AGN is, by definition, a nuclear object, we looked
for good correspondence between these reference
positions and {\it individual} X-ray sources.   With
the exception of NGC 891, we clearly identified X-ray counterparts for
all sources, shown by the 
circular regions in \fr{fig:xo}.
Background regions, with sufficient numbers of counts, and
sources masked out, were also selected.  For sources surrounded by
diffuse emission, this was included in the background. In two cases,
there was no clear best (closest) candidate X-ray
counterpart and we chose two sources, marked by U and L
\footnote{The \lq\lq upper source\rq\rq, U, is at higher
declination than the \lq\lq lower source\rq\rq, L.} in
\tr{tab:sample}. Details of the X-ray/optical sample, including values
for calculated quantities, are given in \tr{tab:sample}.

We calculated \lx\ by using the distances given by
\citet{1997ApJS..112..315H} and flux values \fx$(0.2-10.0 \, {\rm keV})$
estimated by using the tool
PIMMS, \citep[v3.8a,][]{1993Legac...3...21M}.
For the
latter task we used the net count rate, a power law spectrum with
index $\Gamma=1.8$ and the Galactic neutral hydrogen column density
\nhgal\ in the direction of the source \citep{1990ARA&A..28..215D}.
As can be seen in \tr{tab:sample}, 
the brightest source has $\log L_X = 40.42$ (\lunits), so if there is a
hidden AGN, it is weak.

\x\ colours can help discriminate source types, especially
when no spectra are available. We calculated the net counts
in three bands: soft ($S\equiv 0.3-1$ keV),
medium ($M\equiv 1-2$ keV), and hard ($H\equiv 2-8$ keV).
Following \citet{2003ApJ...595..719P},
we define H1 $\equiv(M-S)/T$ and H2 $\equiv (H-M)/T$ to be the soft and hard
colours, respectively. 
T represents the total counts in all three bands combined.
The \x\ colours for each source appear in \tr{tab:sample}.
For sources with no spectra, the calculated colours are
shown in the colour-colour plot of \fr{fig:prest}. 
Regions corresponding
to different source types, taken from \citet{2003ApJ...595..719P},
are marked, as explained in the figure caption.

Finally, we calculated circum-nuclear star formation rates (SFRs),
using the \ha\ luminosities given in HFS and the relation of
\citet{1998ARA&A..36..189K}. Results are presented in \tr{tab:sample}.
The purpose of this was to investigate whether there is any link
between high SFRs and AGN activity, as the gas fuelling star formation
may also sustain an AGN.


\subsection{X-ray spectra}
For all individual X-ray sources identified and having
$>60$ net counts we extracted spectra by
using the CIAO 3.2 script {\tt psextract}. Four spectra were extracted
in all, the faintest extracted source having 333 counts. The spectra were
binned to have 15 counts per bin and analysed with the XSPEC
software package, version 12.2.0.
Details of the spectral results
are given in \tr{tab:spec}. All spectra were fitted with
models which included at least a power law and Galactic absorption.
There is a significant range in spectral shapes (see below).
Any intrinsic
absorption appears to be no more than a few $\times 10^{21}$\cunits.
At low energies a thermal component is needed to account for
a pronounced spectral \lq bump\rq.

In two spectra there is clear indication of line emission between 6
and 7 keV.  It is common practice to estimate the significance of
including an emission line as an additive component to the spectral
model by means of a standard $F$-test.  However, this can be
misleading. As explained by \citet{2002ApJ...571..545P}, one of the
conditions for the use of $F$-statistics stipulates that the null
values of additional parameters introduced may not be on the boundary
of the set of possible parameter values. As the line flux must be
non-negative, its null value is zero, which is the boundary of
non-negative numbers.  In such cases, one has to modify the standard
procedure by performing a {\it calibration} of the $F$-test applied on
the observed spectrum ($F$-test calibration, FC).  Briefly, one
carries out Monte Carlo simulations of spectra using the parameters
for the best {\it null} model fit, i.e. without an emission line, and
the same instrument response, background, exposure time and binning as
for the observed spectrum. Each simulated spectrum is fitted twice,
once with a model that includes the emission line, and once with a
model without the emission line. The $F$-test is applied to each
fitting pair, and the ensemble of such $F$-test results is compared to
the single $F$-statistic value and associated probability result from
the pair of fits to the {\it observed} spectrum, thus providing an
estimate of the reliability of the calculated significance of the
observed $F$-statistic.  In our case, we applied this method by
carrying out 2,000 Monte Carlo simulations for each observed spectrum.

We now discuss each case in greater detail:
\subsubsection{NGC 891}
Optical imaging shows that this source has an edge-on
orientation. This may be one reason why, as can be seen in
\fr{fig:xo}, there is no candidate X-ray counterpart in the ACIS field
corresponding to this source. Accordingly, the flux and luminosity
values in \tr{tab:sample} are upper limits, and we do not discuss this
source any further.

\subsubsection{NGC 4303}
The isolated X-ray source shows 728 net counts. The centre of the
selected circular source region is at $\ga 3$\arcsec\ from the optical
position. Binning of the X-ray image reveals that the X-ray/optical
pair appears roughly in the centre of a ring of diffuse X-ray
emission. HFS classify this galaxy as an \htwo\ nucleus, although
their values for \ntwof/\ha\ and \stwof/\ha\ favour an AGN
classification.  This is because HFS give less weight to the
\ntwof/\ha\ ratio due to the possibility of selective nitrogen
enhancement in galactic nuclei, and the \stwof/\ha\ value is
borderline between \htwo\ and AGN.

The X-ray spectrum (\fr{fig:4303spec}) is best fitted with a
redshifted power-law plus hot diffuse gas
model (MEKAL), with only the default
parameters kept fixed.  There is
no indication of emission in the spectral region around 6.4 keV.


\subsubsection{NGC 3310}
More than ten discrete X-ray sources, embedded in diffuse emission,
can be seen in the ACIS image of this galaxy. One of the sources
corresponds extremely well with the optical centre, and has been
isolated for further study here.  As this isolated source shows 1538
net counts, a high-quality spectrum can be extracted.  Up to $\sim 7$
keV, the continuum is well defined, making a prominent emission
feature clearly noticeable at 6.4 keV at a level of $\sim
2\sigma$ above the local continuum. The spectrum (\fr{fig:3310spec})
is well fitted with a power law, a thermal component and moderate
intrinsic absorption.  
The power law is flatter than the canonical average for AGN.
Fixing $\Gamma$ at a canonical AGN value of 1.8 leads to a somewhat
worse \cs, but the intrinsic column density remains virtually unchanged.

The best fit (reduced \cs~$\equiv$~\csn~$\sim 0.8$) includes a
gaussian component for an emission line fixed at $\sim 6.4$ keV with
an equivalent width (EW) of 0.3 keV. A similar, if slightly poorer, fit is produced if the
line energy and Gaussian $\sigma$ of the line profile are left free.
In this case, the best fit energy is $6.4\pm0.1$ keV. This
cannot be achieved with a 6.7 keV line. Both the energy and Gaussian $\sigma$
must then be fixed, but this produces a poorer fit 
than without an emission
line at all, and thus no $F$-test can be applied for such a line. 
Plotting data and fits shows clearly that only a narrow line at
6.4 keV is able to fit the sharp feature at $> 6$ keV above the
well defined continuum. 

Indeed, the calibrated $F$-test suggests that the fitted
emission line is statistically significant: The $F$-statistic value
for the observed spectrum is 4.8, with random probability 0.03. FC
shows that more than 99\% of the simulated spectra lead to a smaller
$F$-value, and, correspondingly, higher random probabilities.


\subsubsection{NGC 5905}
This is the most distant source in the sample. The level of net counts
(40) is low, precluding the extraction of a useful spectrum.  
Both \x\ colours are quite soft, suggesting a source with
$\Gamma\sim 2.5$. In the colour-colour
diagram of \fr{fig:prest} this source is located
in the low-mass \x\ binary (LMXB)
region.

\subsubsection{NGC 2782}
The X-ray source 
has 597 net counts. It can be seen in \fr{fig:2782spec} that the
extracted spectrum has a pronounced bump at energies below $\sim 1.5$
keV which is well fitted by a hot diffuse-gas model. The high energy
region is fitted by a power law which appears to be flat. The best fit
(\csn~$=1.0$) requires a gaussian component at 6.4 keV, (EW $\sim 1.5$
keV) with a high-significance $F$-statistic value of 17.0 and
associated random probability $3 \cdot 10^{-4}$.  FC shows that all
2000 simulated spectra lead to smaller $F$-values.

As in the case of NGC 3310, if the line energy is left free,
XSPEC places the line at $6.4\pm0.1$ keV.
Fitting a line at 6.7, rather than 6.4, keV requires fixing all
line components, the fit is clearly poorer (\csn~$=1.4$) and,
as the number of degrees of freedom
is the same as without a line, the
$F$-test is inapplicable.

\subsubsection{NGC 4102}
This is a clear, isolated X-ray source with sub-arcsecond separation
from its optical counterpart, and 333 net counts. The extracted X-ray
{spectrum is shown in \fr{fig:4102spec}.  A power-law fit provides a
mediocre fit, which is somewhat improved by the addition of a
Raymond-Smith hot, diffuse-gas model ($F$-test probability 0.1). At
energies $\ga 5$ keV the continuum appears to be sloping upwards. This
feature can be fitted by a Gaussian at $\sim 9$ keV with an EW $\ga
450 $ keV. Both values are not well constrained but lead to a much
improved fit. To test the reality of this feature, 
we used several, distinct background regions 
and performed a separate spectral extraction,
but this feature persists. At a lower
grouping of 5 counts per bin this feature looks more
like a noisy broad bump around 7 keV with an equivalent width
$\sim 200$ keV. However, this line cannot be shown to be
statistically significant.
We thus do not take this feature to be indicative
of \feka\ or $\beta$ emission near 7 keV, although, the excellent
agreement between the optical and X-ray positions would make this
object a good hidden AGN candidate.

\subsubsection{NGC 4217}
There are two isolated X-ray sources, U and L, closest to the optical
position. These have 21 and 9 counts, respectively, which are very low
for a useful spectrum to be extracted.  Additionally, the flux and
luminosity information for source L are $2\sigma$ upper limits.  
Both sources fall in the the high-mass \x\ binary (HMXB) region of 
\fr{fig:prest}.
Their \x\ colours are hard and
are suggestive of a relatively flat photon index ($\Gamma\sim 1$) and high
intrinsic absorption (few $\times 10^{22}$\cunits).

\subsubsection{NGC 2146}
As in the case of NGC 3310, the ACIS image of this galaxy shows
several discrete point sources embedded in diffuse emission. The
source that best corresponds to the optical position has only 32 net
counts, but the positional agreement is the best among all of our
X-ray/optical pairs. 
Given its \x\ colours, this source may have an intrinsic $\Gamma \sim 1.8$
and \nh~$\sim$~\ten{4}{22}\cunits, consistent with an AGN with high intrinsic
absorption. Alternatively, the source is also consistent with
a flat $\Gamma \sim 0.3$, \nh~$\sim$~\ten{1}{22}\cunits, and falls in the
HMXB region of \fr{fig:prest}.

\subsubsection{NGC 278}
As in the case of NGC 4217, there are two candidate X-ray
counterparts, U and L, roughly equidistant from the optical position,
with 60 and 29 counts, respectively.  
The \x\ colours for both sources are quite soft, suggesting
no intrinsic absorption, steep photon indices ($2.3-2.7$), and
are consistent with LMXBs.

\section{Discussion}
\subsection{NGC 2782 and 3310 revisited}
There is only a handful of star-forming galaxies known to harbour a
LLAGN, revealed at X-ray wavelengths. Out of nine HFS sources which
fall within ACIS-S fields, we extracted four spectra. In two out of
these, we detect a 6.4 keV \feka\ line with high significance.
In contrast, the data for these isolated, central
sources do not corroborate a 6.7 keV emission line, often associated with
star-formation \citep{2002Natur.415..148W,2005ASPC..331..345S}.
Due to the small number of relatively early-type \htwo\ nuclei from
HFS, whose positions have been observed in the X-ray band, the
significance of the detection fraction (2 out of 9) remains, of
course, unclear.

The NGC 2782 detection has the highest significance although the
spectrum is noisier, the continuum less well-defined, the model
goodness-of-fit and the optical/X-ray positional correspondence
poorer.  In spite of these caveats, the isolated central X-ray source
has the highest \lx\ in our sample.  Furthermore, the detection of a
hidden AGN in this object would not come as a surprise.  Within the
context of \rosat\ observations, the possibility of a LLAGN was
already considered by \citet{1998A&A...330..823S} and the galaxy is in
fact classified as a Seyfert 2 by \citet{2003A&A...412..399V}. The
galaxy is number 215 in Arp's Atlas of Peculiar Galaxies
\citep{1966ApJS...14....1A} and has Hubble type classification
SAB(rs)a pec \citep{2005ApJ...618L..21W}. Amongst nearby spirals, it
harbours one of the most powerful, M~82-like, nuclear starbursts
\citep{1983ApJ...268..602B,1989ApJ...346..126D}, fuelled by a massive,
gas-rich nuclear bar, driving molecular gas towards the central
regions at large gas-inflow rates, $> 1$ \msuny\
\citep{1999ApJ...526..665J}.  A pair of prominent \hone\ and optical
tails surrounding an optical disk with three ripples
\citep{1991ApJ...378...39S,1999AJ....117.1237S} suggest that the
starburst may have been triggered by an interaction or merger
\citep{1994AJ....107.1695S}. The presence of large amounts of gas and
an efficient fuelling mechanism indicate that there should be plenty
of material both for feeding an AGN co-existing with the powerful
starburst and for absorbing soft X-rays so as to produce the hard,
flat power-law and a \feka\ line with a moderately high EW $\ga 1$ keV.
Indeed, the \ha-calculated SFR for this source is the highest among all sources in
our sample (\tr{tab:sample}).

\citet{2002ApJ...573L..81L} have shown that \feka\ equivalent widths
correlate with geometry: Larger EWs imply an \lq AGN-Type 2\rq\ view, whilst
smaller EWs an \lq AGN-Type 1\rq\ view, in which the \x\
continuum is viewed directly.
Although, their models are for heavily obscured,
Compton thick AGN, high intrinsic column densities are certainly
plausible in this galaxy. 

%

The third highest \lx\ in our sample is for NGC 3310. This galaxy (Arp
217), is also an M~82-class starburst, of type SAB(r)bc pec
\citep{1991trcb.book.....D}. Its peculiar optical morphology and tidal
debris \citep{2005ApJ...618L..21W} suggest that the starburst has been
triggered by a merger with either a gas-rich dwarf companion
\citep{1981A&A....96..271B,1996ApJ...473L..21S} or a similar-size disk
\citep{2001A&A...376...59K}.  The nuclear spectrum has been explained
by using only photoionisation by hot stars
\citep{1993RMxAA..26..124P}. Until now there has been no indication of
AGN activity in any wavelength band. In particular, work with \rosat\
\citep{1998MNRAS.301..915Z} and \xmm\ \citep{2004MNRAS.352.1335J} has
found no indication of X-ray emission due to an AGN in this
object. However, \rosat\ has lower sensitivity and resolution than
\chandra, whilst \citet{2004MNRAS.352.1335J} have excluded the energy
range $> 6$ keV in the analysis of the \xmm\ data.  It would then be
the first time that this object can be searched for \feka\ emission,
and the evidence suggests that such emission has been detected.
The \ha-calculated star-formation rate for this source, is the fourth
highest in our sample.

We calculated the black hole mass, \mbh, and the corresponding
Eddington luminosity, \ledd, in these two galaxies by using velocity
dispersion, $\sigma$, values from \citet{1995ApJS..100..105M} and the
\mbh-$\sigma$ relation of \citet[][Equ.~20]{2005SSRv..116..523F}.  We
estimated the bolometric luminosity of the isolated, central sources,
assuming \lbol~$\approx 10$~\lx\ \citep{2003ApJ...589L..13N}.  This
gives \lbol/\ledd$\sim$~\ten{6}{-5} and \ten{9}{-5} for NGC 2782 and NGC
3310, respectively. Thus, the luminosity due to black hole accretion
is significantly sub-Eddington.  Furthermore, the fraction of
X-ray flux coming from the central source, to that from the whole
galaxy is $\sim 20$\%\ and 10\%\ for NGC 2782 and NGC 3310, 
respectively. It follows, that any AGN contribution to the energy
output of these objects is small.

\subsection{The case of NGC 4303}
We believe the case for a hidden AGN in this galaxy is strong,
although it cannot be corroborated by detection of Fe emission.  The
galaxy has morphological type SAB(rs)bc \citep{1991trcb.book.....D}
and is classified as Seyfert 2 by
\citet{2003A&A...412..399V}. \citet{2000ApJ...529..845C} and
\citet{1997ApJ...484L..41C} also label the galaxy as \lq active\rq. It
has a large-scale, outer bar \citep{2002ApJ...567...97L}, as well as a
smaller inner bar, surrounded by a $\sim 0.5$ kpc circum-nuclear
starburst ring
\citep{2000ApJ...529..845C,2000MNRAS.317..234P,1997ApJ...484L..41C}.
Using \rosat\ observations, \citet{2000A&A...360..447T} have detected
several point sources across the face of this spiral which they
attribute to HMXBs, tracing actively star-forming \htwo\ regions. In
\fr{fig:4303con} we overplot contours from the binned X-ray image on
an \ha\ image of NGC 4303 \citep{2004A&A...426.1135K}. As can be seen
from this figure, as well as Fig.~1 of \citet{2003ApJ...593..127J},
the spiral pattern is traced well in the X-rays.  The centermost
contours in \fr{fig:4303con} most likely trace the galaxy's
circum-nuclear ring which has been resolved with \hst\ in the
ultra-violet band (UV) by \citet{1997ApJ...484L..41C} and studied
extensively by \citet{2003ApJ...593..127J}.  In spite of detailed
evolutionary synthesis modelling, these authors cannot unambiguously
distinguish among different possibilities for explaining the central
X-ray emission, including a number of hidden AGN scenarios, a central
star cluster, or Ultra-Luminous X-ray (ULX) sources.

\subsection{The rest of the sources}
For the rest of the sources, we do not obtain any strong evidence for
AGN activity. \lx\ values are not high enough to independently suggest
AGN activity, but this is expected as we are looking for hidden
AGN. The lack of adequate spectra makes it impossible to search for
\feka\ emission. We only get a hint as to the nature of these sources
by searching $\Gamma-$~\nhint\ space for pairs of values that will
reproduce the \x\ colours, H1 and H2, calculated from the image.  The
location of the sources in the H1$-$H2 plane (\fr{fig:prest}) provides
some insight as to their possible XRB nature.  Otherwise, since neither
$\Gamma$ nor \nhint\ can be established independently, this procedure
cannot be used to claim detection of a hidden AGN.  In any case, this
technique suggests that one of these sources is consistent with an AGN
with high intrinsic absorption. {\it All} sources are consistent
either with low- or high-mass XRBs.  X-ray spectra would be needed to
further explore these possibilities. Such spectra would further allow
to differentiate between ULXs and AGN. However, a defining
characteristic of ULXs is their off-nuclear nature, and all of these
sources appear to be nuclear.

\subsection{The starburst-AGN connection}
There appears to be an intimate link between starbursts and AGN which
is not yet well understood. One the one hand, high \lx\ in composite
galaxies suggests an AGN, although based on their optical spectra
these are classified, at best, as being on the border between AGN and
star-forming galaxies. On the other hand, we detect LLAGN in bona
fid{\ae} star-forming galaxies. All these objects have one common
trait: they include a powerful starburst, indicative of a large gas
supply to feed a central engine. In our sample,
the most significant detection of 6.4 keV emission comes
from the source with highest \ha-calculated SFR.
A disturbed morphology, including
bars and recent or on-going interactions are further clues to an
efficient feeding mechanism.

This does not imply that {\it all} such cases will at present accrete
onto a super-massive black hole. In the sample of
\citet{2001AJ....121..662B} luminous hard X-ray sources are common in
bulge-dominated optically luminous galaxies, with about 10\% of the
population showing activity at any given time.  These authors suggest that this
fraction may be interpreted in terms of a duty cycle of galaxies,
which are active at any given time.

Alternatively, an evolutionary scheme such as the one proposed by
\citet{2004ApJ...605..144M} will conceivably limit the number of
sources that actually have an active component at any given time,
irrespective of power. We may then simply be detecting galaxies which,
according to the terminology of these authors, are in a \lq starburst
dominant\rq\ or \lq starburst-dominant Seyfert\rq\ phase, whereas
galaxies lacking an AGN signature may be in an \lq inactive\rq\
phase. Put differently, there is a continuum of galactic \lq
activity\rq\ levels, and the level of the galaxies with a hidden AGN
detected in our work is lower than the galaxies of
\citet{2001ApJ...558...81C}. In the case of NGC 253, which like our
sources, also has low \lx\ and line emission suggesting a buried AGN,
\citet{2002ApJ...576L..19W} suggest that one may be detecting the
ending or beginning of AGN activity.

Although it might be instructive to do so, we have not attempted to
use infra-red diagnostics of AGN/star-forming activity, such as
$16\mu/24\mu$ or $70\mu/24\mu$ flux ratios. However, given the low
level of AGN activity in these objects, it would be surprising if such
tools were to provide any further insight.



%

\section{Summary}  
We used \chandra\ data to look for X-ray signatures of AGN in an
optically selected sample of star-forming galaxies from HFS. We
identified and isolated 10 individual X-ray counterparts to optical
nuclear sources. We produced X-ray spectra for four of these, and
used \x\ colours to explore $\Gamma-$\nhint\ space for the rest. 
We estimated SFRs, using \ha\ luminosities. 
Our main conclusions are:
\begin{enumerate}
\item All sources have low \lx, not indicative of powerful AGN.
\item For the strong starbursts NGC 2782 and NGC 3310, we
detect a 6.4 keV \feka\ line with high significance.
However, these AGN have significantly sub-Eddington luminosities,
whilst the level of contamination of the total galactic
X-ray flux is small. NGC 2782 has the highest SFR, and NGC 3310
the fourth highest in our sample.
\item In spite of the lack of \feka\ line
emission, the powerful starburst NGC 4303 is a good candidate for harbouring
a hidden AGN.  For sources with no spectra, \x\ colours are consistent
with LMXBs (three sources), HMXBs (three sources), 
or a hidden AGN (one source).
\end{enumerate}

The detection of underlying AGN activity in apparently \lq pure\rq\
star-forming galaxies is a further indication that the two components often
co-exist and that there is no clear cut-off below which any form of
AGN activity can be ruled out altogether.

\section{Acknowledgements} 
This work was funded in part by the Hellenic General Secretariat for
Research and Technology within the framework of the Hellas-USA
collaboration programme {\it Study of Galaxies with the Chandra X-ray
Satellite}.  This research has made use of data obtained from the High
Energy Astrophysics Science Archive Research Center (HEASARC),
provided by NASA's Goddard Space Flight Center. This research has made
use of the NASA/IPAC Extragalactic Database (NED) which is operated by
the Jet Propulsion Laboratory, California Institute of Technology,
under contract with the National Aeronautics and Space Administration.

\tiny
\newcommand{\noopsort}[1]{}

\begin{figure*}
\centering

\rotatebox{270}{\includegraphics[width=6cm,angle=90]{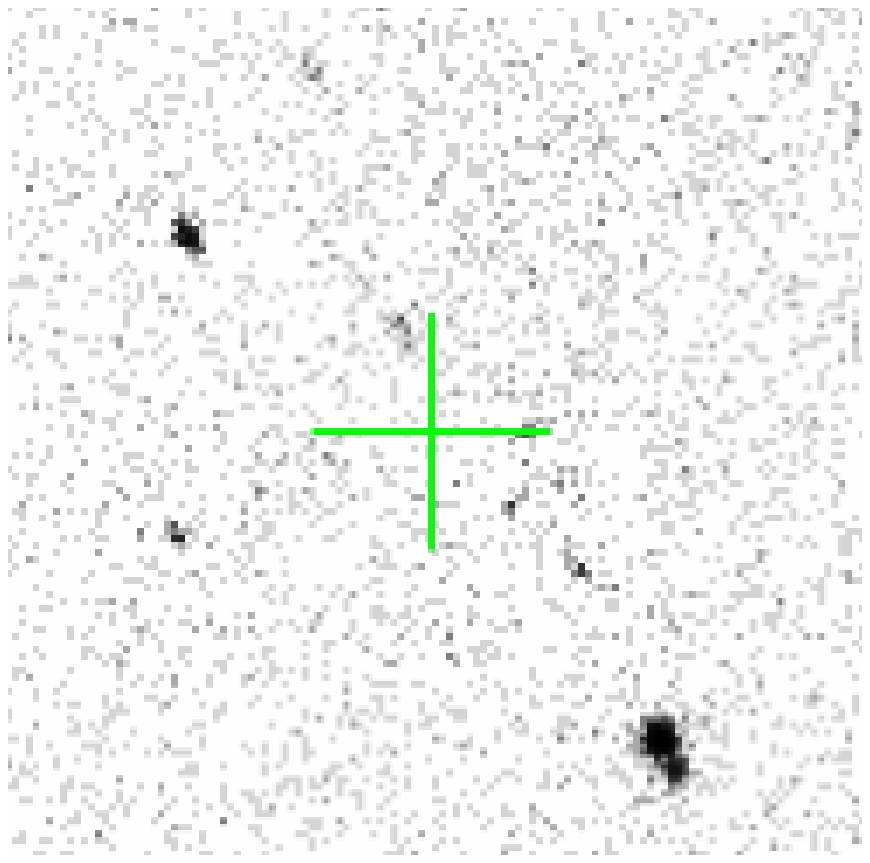}}
\hspace{1cm}
\rotatebox{270}{\includegraphics[width=6cm,angle=90]{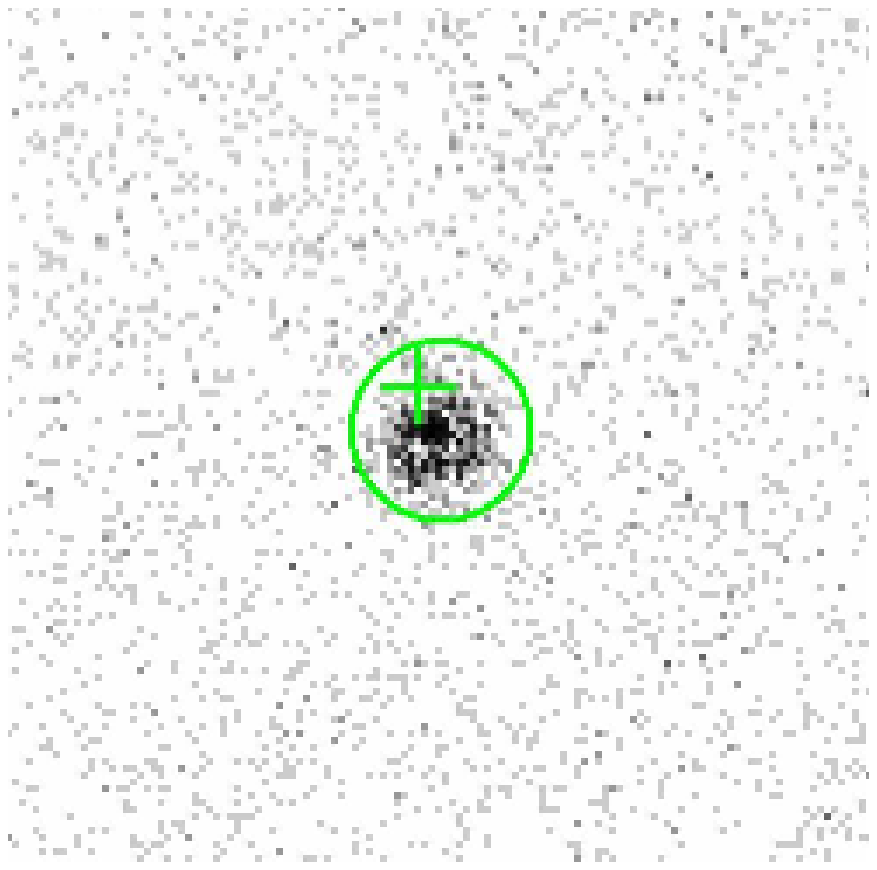}}\hfill \\

\vspace{1cm}
\rotatebox{270}{\includegraphics[width=6cm,angle=90]{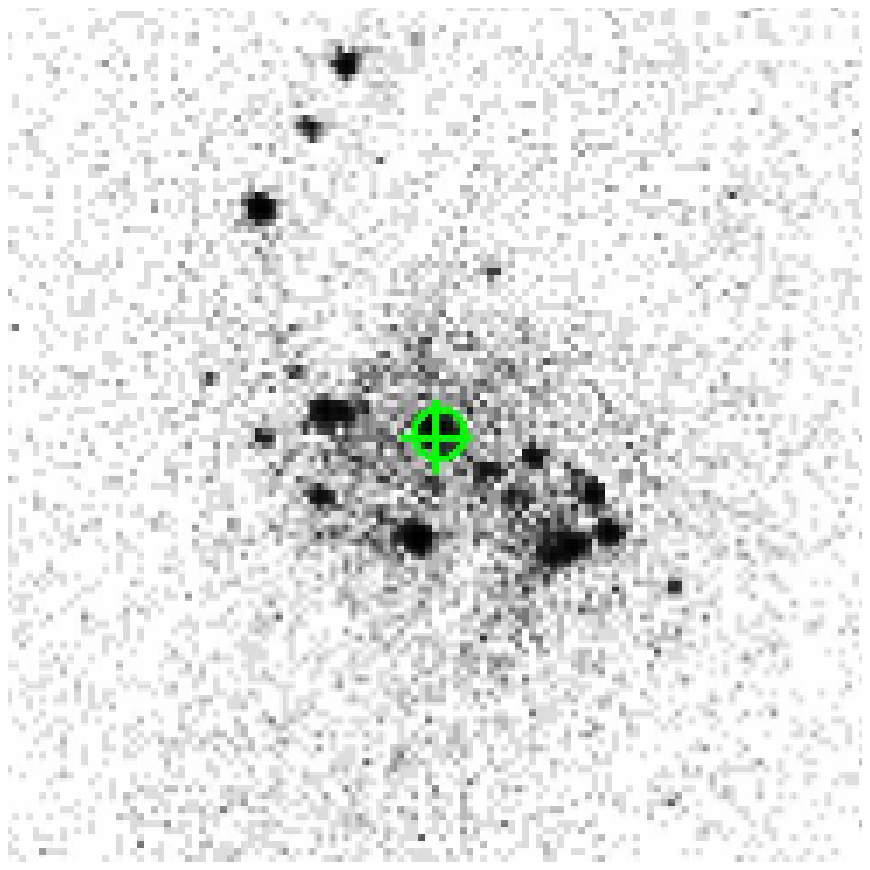}}
\hspace{1cm}
\rotatebox{270}{\includegraphics[width=6cm,angle=90]{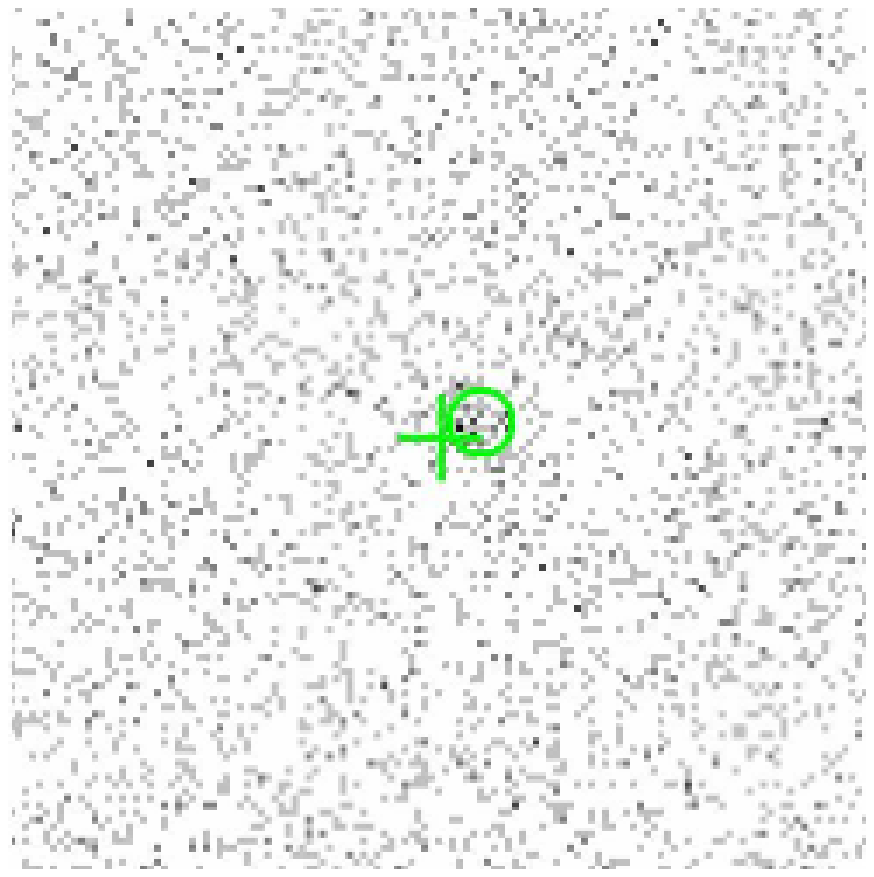}}\hfill \\

\vspace{1cm}
\rotatebox{270}{\includegraphics[width=6cm,angle=90]{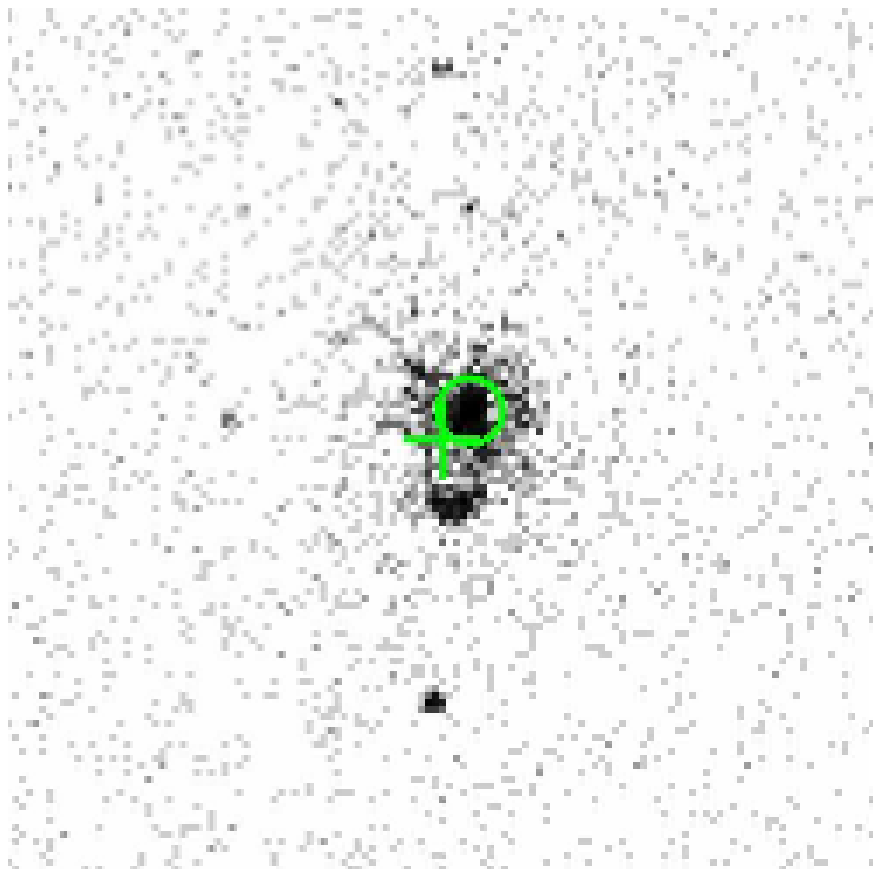}}
\hspace{1cm}
\rotatebox{270}{\includegraphics[width=6cm,angle=90]{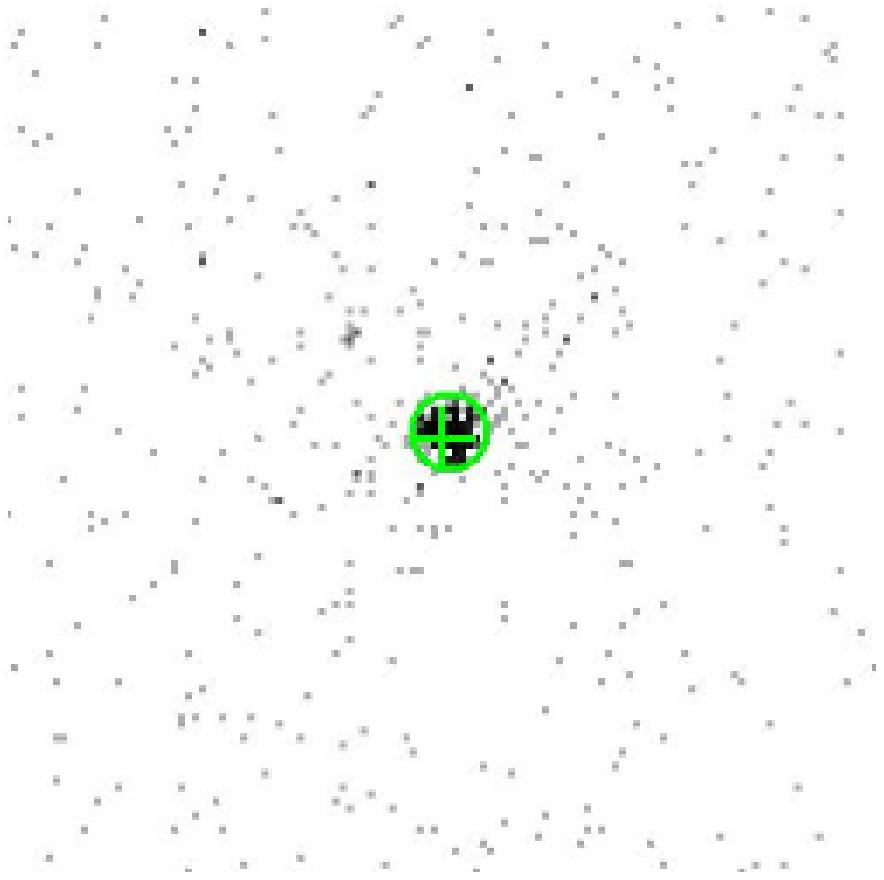}}\hfill \\

\caption{\chandra/ACIS-S images of galaxies in this project.  North is
up and east is to the left. From left to right and top to bottom the
galaxies shown are NGC 891, NGC 4303, NGC 3310, NGC 5905, NGC 2782,
NGC 4102. The cross and its size represent the best optical position
and its uncertainty from \citet{1999ApJS..125..409C}. The circular
region nearest to the cross represents the best candidate for an X-ray
counterpart. All images are unbinned and have a size of
61\arcsec~$\times$~61\arcsec.}
\label{fig:xo}
\end{figure*}

\addtocounter{figure}{-1}
\begin{figure*}
\centering

\rotatebox{270}{\includegraphics[width=6cm,angle=90]{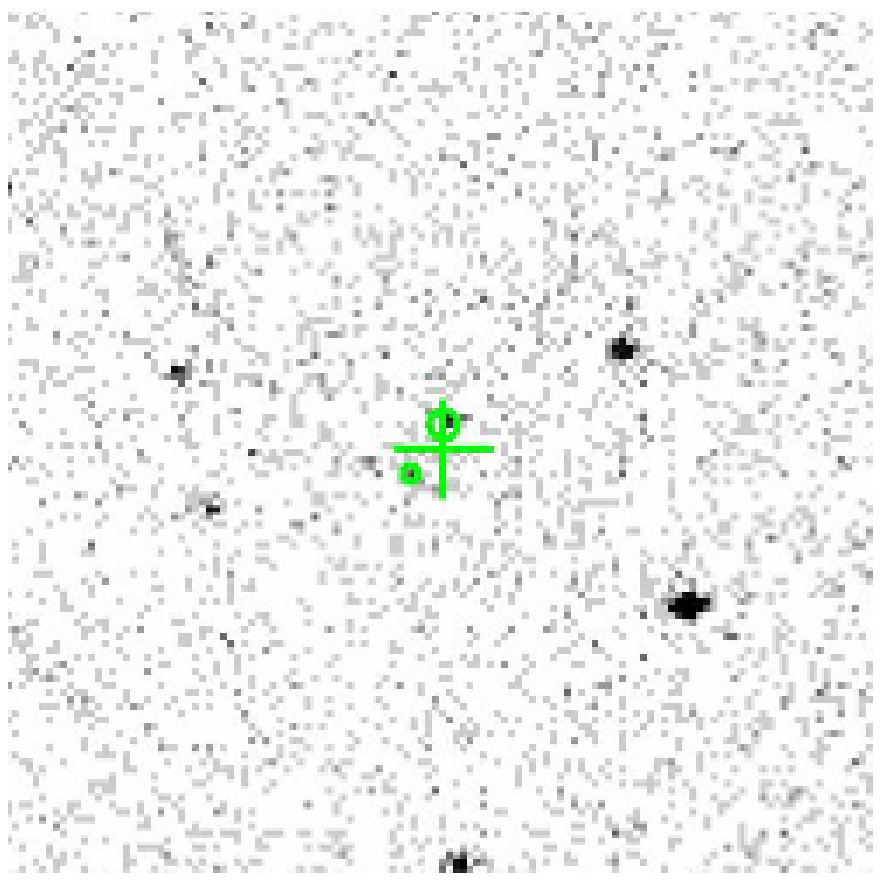}}
\hspace{1cm}
\rotatebox{270}{\includegraphics[width=6cm,angle=90]{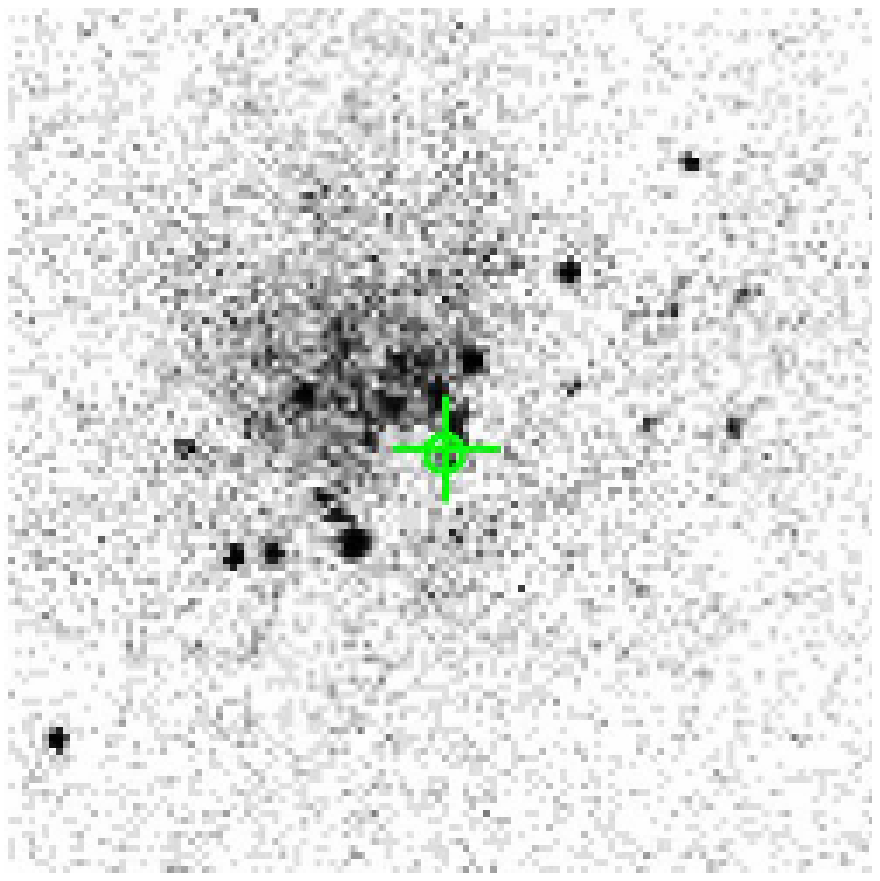}}\hfill \\

\vspace{1cm}
\rotatebox{270}{\includegraphics[width=6cm,angle=90]{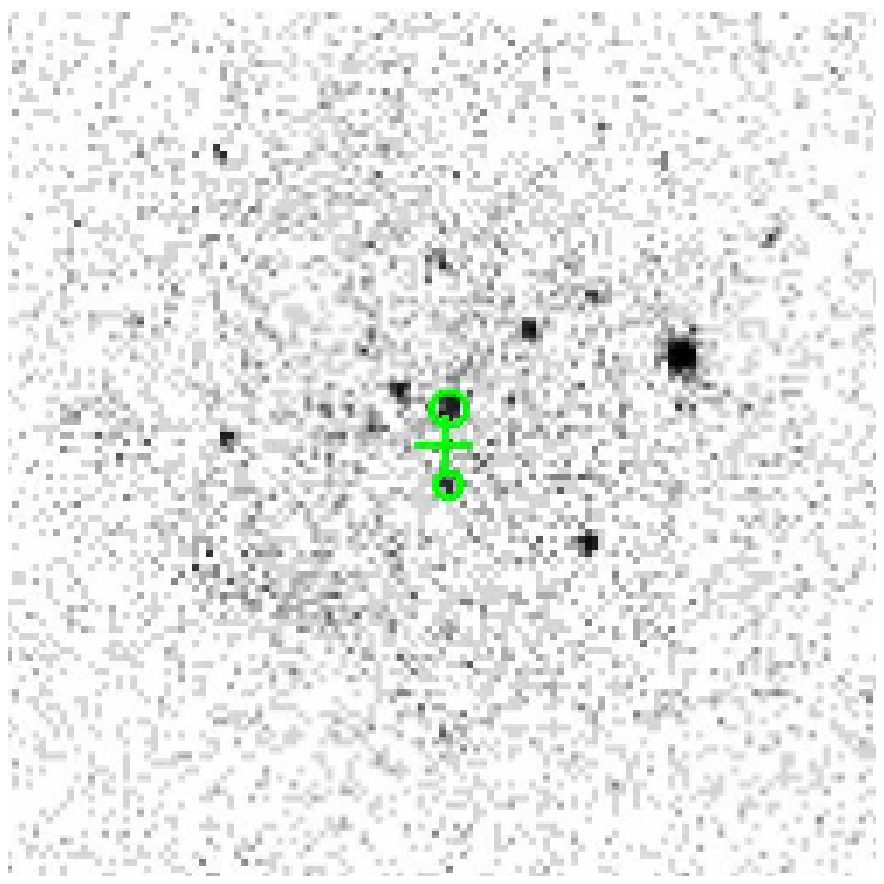}}
\hspace{1cm}
 \\

\caption{{(\it cont.}) Images of NGC 4217, NGC 2146, NGC 278. }

\end{figure*}

\begin{table} 
\scriptsize
\caption{\chandra\ observation log. Columns give (1) target name, (2) \chandra\
sequence number, (3) observation ID, (4) exposure time, (5)
Galactic neutral hydrogen column density. All
observations are with ACIS-S.}
\label{tab:X}
\centering
\begin{tabular}{rc cr r}
\hline\hline
Name    &  Sequence no. & Obs. ID & $t_{\rm exp}$  & \nhgal \\
        &               &         &   (ks)         & ($10^{20}$\cunits) \\
\hline
NGC 278  &  600189 & 2055 & 38.3 & 13.3 \\
NGC 278  &  600190 & 2056 & 37.3 &      \\
\hline
NGC 891  &  600097 &  794 & 50.9 & 7.64 \\ 
\hline
NGC 2146 &  700570 & 3131 &  9.3 & 7.30 \\
NGC 2146 &  700571 & 3132 &  9.5 &      \\
NGC 2146 &  700572 & 3133 &  9.8 &      \\
NGC 2146 &  700573 & 3134 &  9.6 &     \\
NGC 2146 &  700574 & 3135 & 10.0 &      \\
NGC 2146 &  700575 & 3136 &  9.8 &      \\
\hline
NGC 2782 &  700453 & 3014 & 29.6 & 1.76 \\
NGC 3310 &  600276 & 2939 & 47.2 & 1.13 \\
NGC 4102 &  700693 & 4014 &  4.9 & 1.77 \\
NGC 4217 &  600393 & 4738 & 72.7 & 1.23 \\
NGC 4303 &  700339 & 2149 & 28.0 & 1.67 \\
NGC 5905 &  700445 & 3006 &  9.6 & 1.44 \\
\hline
\hline
\end{tabular} 
\end{table}

\begin{table*} 
\scriptsize
\caption{HFS and \chandra\ galaxies. These are relatively early-type galaxies
classified as \htwo\ nuclei by HFS, which have also been observed with
\chandra. Shown from left to right are (1) sample identification number,
(2) name, (3) right ascension and (4) declination for the X-ray source,
(5) offset between X-ray and optical \citep{1999ApJS..125..409C} position,
(6) $B$ magnitude from HFS, (7) X-ray flux, (8) redshift, (9) logarithmic
X-ray luminosity, (10) soft X-ray colour, (11) hard X-ray colour, and (12) 
circum-nuclear star formation
rates calculated using \ha\ luminosities taken from HFS and the relation of
\citet{1998ARA&A..36..189K}.} 
\label{tab:sample} 
\centering 
\begin{tabular}{cccc ccrc rr r r} 
\hline\hline 
ID &       Name    & $\alpha_X$     & $\delta_X$  & $\delta_{XO}$ &$B$  & $f_X/10^{-14}$ & $z$   & log $L_X$& H1 & H2 & SFR(\ha) \\ 
   &               & (J2000)        & (J2000)     &   (\arcsec)   &     &(cgs)           &       & (cgs)    &    &    & \msuny\   \\
\hline 
1  & NGC278U  &  00 52 04.4 & +47 33 04   & 2.62 & 11.5 & 0.909   & 0.0028 & 38.18 & $-0.05$ & $-0.23$ & \ten{4.7}{-3}\\  
2  & NGC278L  &  00 52 04.4 & +47 32 58   & 2.70 & 11.5 & 0.437   & 0.0028 & 37.86 & $0.07$ & $-0.30$  & \\  
3  & NGC891   &  02 22 32.9 & +42 20 45   & 0.49 & 10.9 & $<$3.910 & 0.0023 & $<$38.63 &  $-$ & $-$    & $>$\ten{1.6}{-4}\\  
4  & NGC2146  &  06 18 37.6 & +78 21 20   & 0.29 & 11.2 & 0.547   & 0.0041 & 38.29 & $0.11$ & $0.79$   & \ten{4.6}{-2}\\  
5  & NGC2782  &  09 14 05.1 & +40 06 47   & 1.85 & 12.1 & 15.700  & 0.0090 & 40.42 & $-0.02$ & $0.05$  & \ten{4.9}{-1}\\  
6  & NGC3310  &  10 38 45.8 & +53 30 11   & 0.37 & 11.2 & 26.400  & 0.0045 & 40.04 & $0.38$ & $-0.01$  & \ten{5.6}{-2}\\   
7  & NGC4102  &  12 06 23.0 & +52 42 40   & 0.71 & 12.3 & 55.500  & 0.0041 & 40.28 & $0.06$ & $-0.12$  & \ten{8.3}{-2}\\ 
8  & NGC4217U &  12 15 50.8 & +47 05 32   & 1.74 & 10.9 & 0.231   & 0.0041 & 37.90 & $0.15$  & $0.40$  & \ten{6.7}{-4}\\
9  & NGC4217L &  12 15 51.0 & +47 05 29   & 2.97 & 10.9 & $<$0.192 & 0.0041 & $<$37.82 & $0.12$  & $0.79$ & \\  
10 & NGC4303  &  12 21 54.9 & +04 28 25   & 3.37 & 10.1 & 21.100  & 0.0037 & 39.77 & $-0.48$ & $-0.17$ & \ten{3.1}{-2}  \\
11 & NGC5905  &  15 15 23.3 & +55 31 02   & 3.08 & 12.3 & 3.350   & 0.0107 & 39.90 & $-0.21$ & $-0.27$ & \ten{1.4}{-1}   \\  
\hline 


\end{tabular} 
\end{table*}

\begin{figure}
\includegraphics[width=8cm,angle=-90]{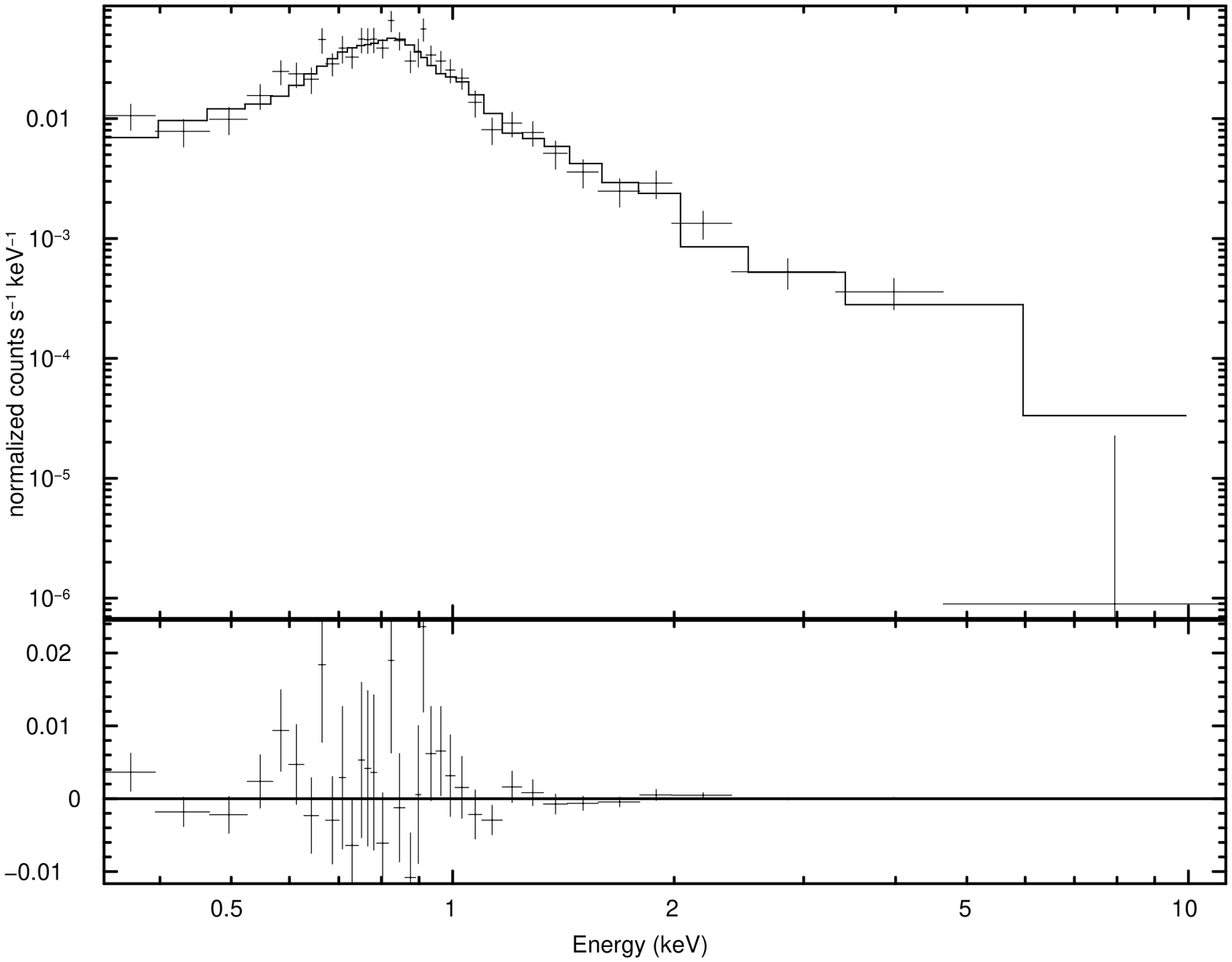}
\caption{Fitted \chandra-ACIS X-ray spectrum of NGC 4303.}
\label{fig:4303spec}
\end{figure}
\begin{figure}
\includegraphics[width=8cm,angle=-90]{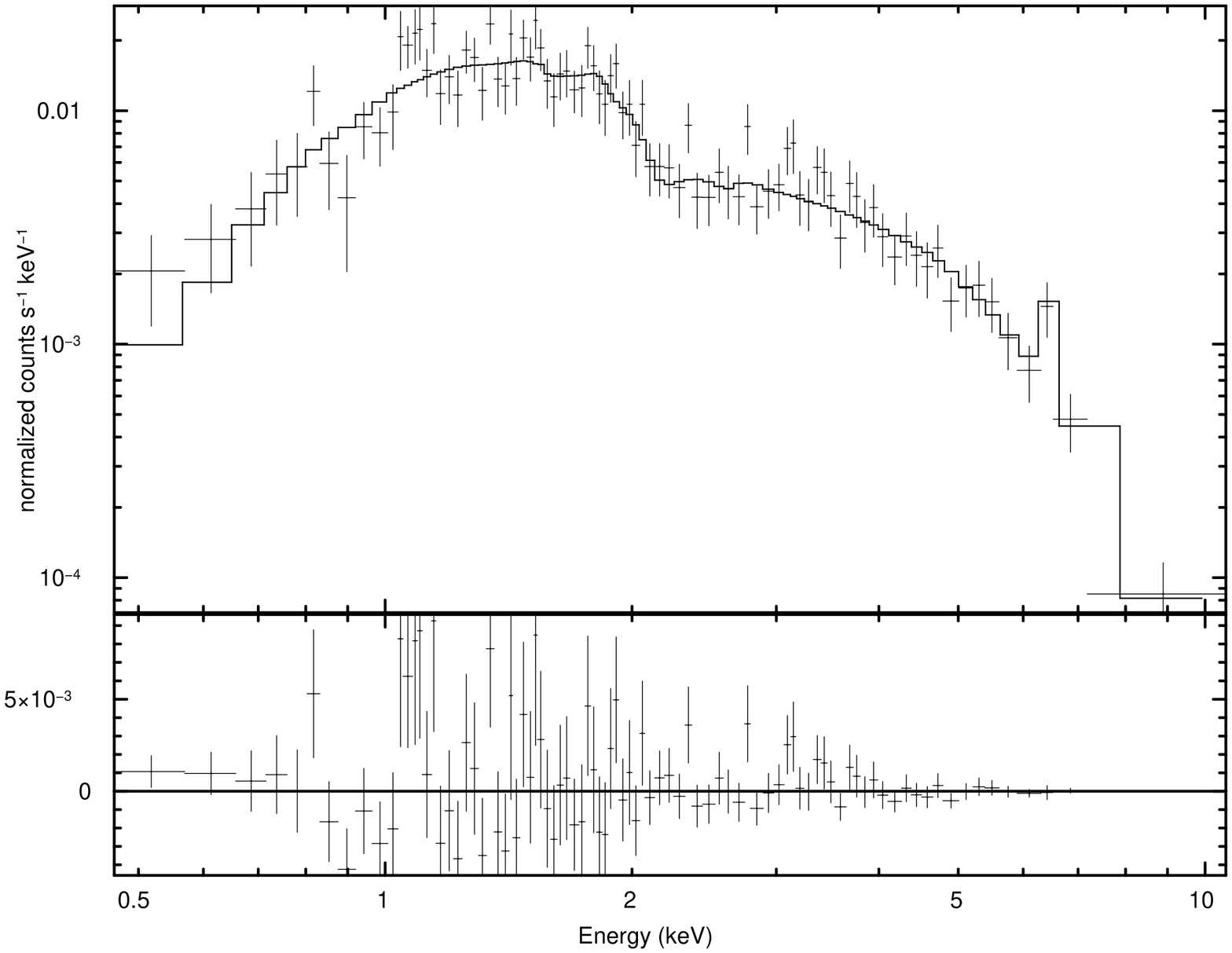}
\caption{Fitted \chandra-ACIS X-ray spectrum of NGC 3310.}
\label{fig:3310spec}
\end{figure}
\begin{figure}
\includegraphics[width=8cm,angle=-90]{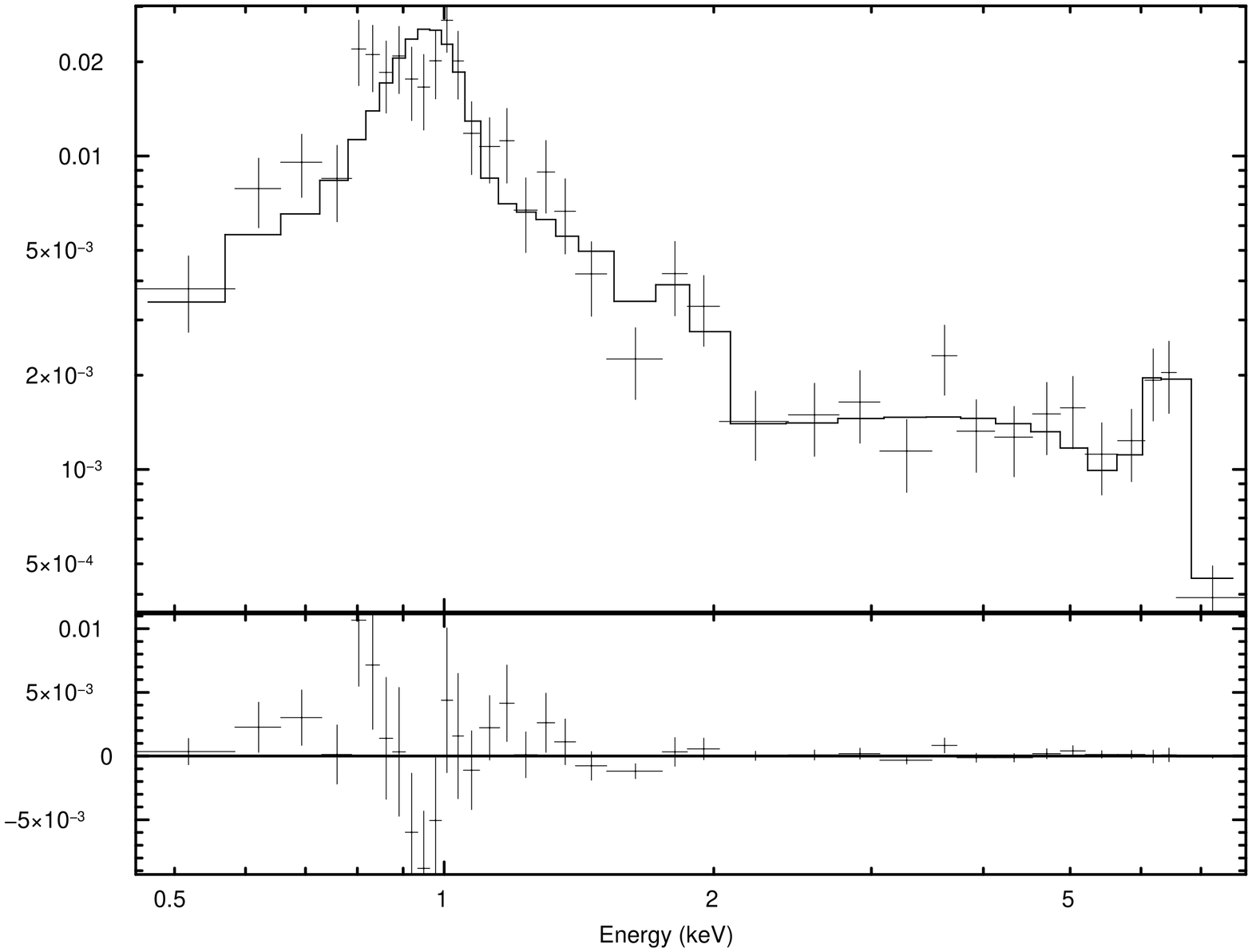}
\caption{Fitted \chandra-ACIS X-ray spectrum of NGC 2782.}
\label{fig:2782spec}
\end{figure}
\begin{figure}
\includegraphics[width=8cm,angle=-90]{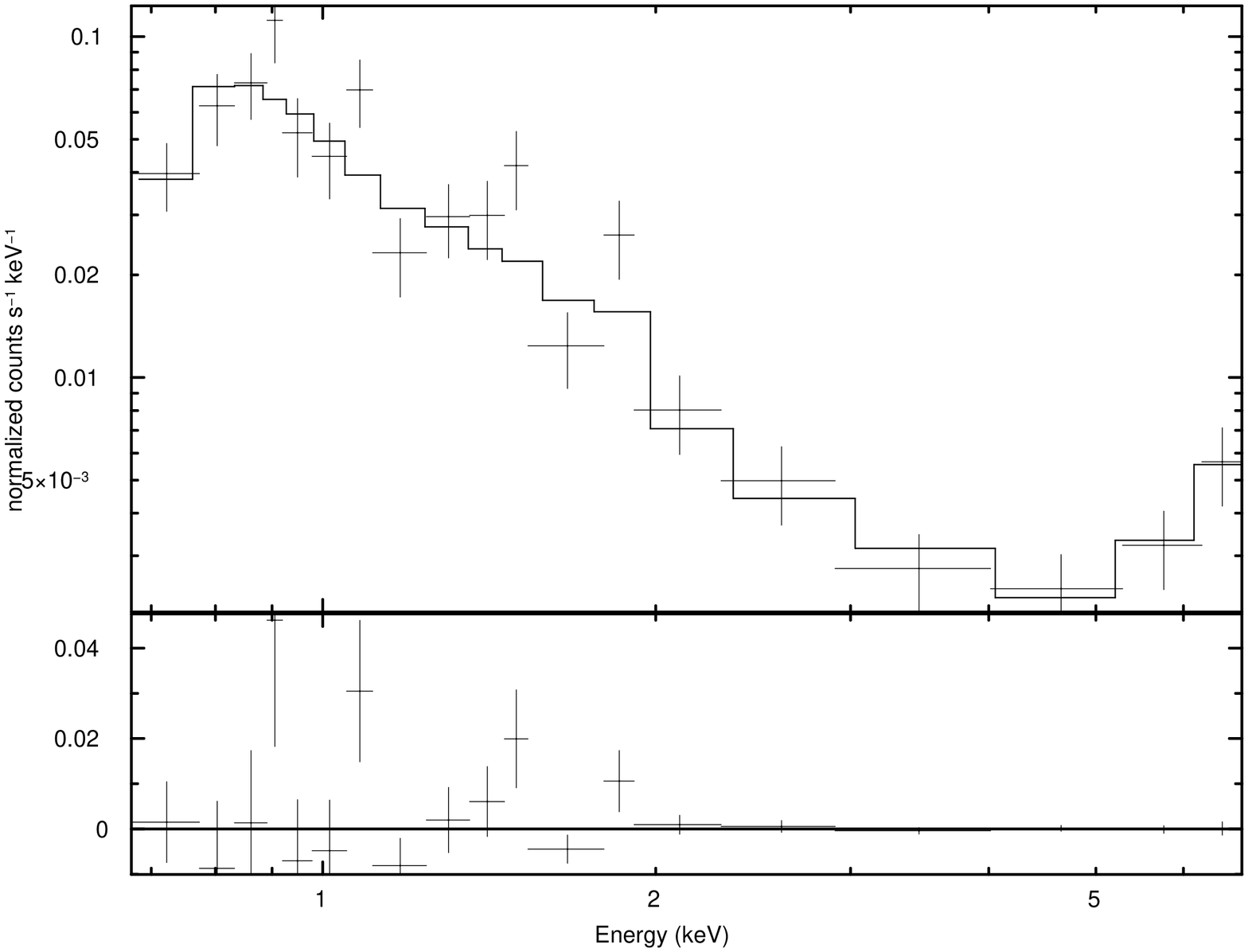}
\caption{Fitted \chandra-ACIS X-ray spectrum of NGC 4102.}
\label{fig:4102spec}
\end{figure}

\begin{figure}
\includegraphics[width=8cm,angle=0]{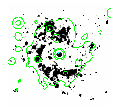}
\caption{X-ray contours for \chandra-ACIS X-ray image of
NGC 4303 overlayed
on \ha\ image \citep{2004A&A...426.1135K} covering 
$\sim $~240\arcsec~$\times$~240\arcsec. The contours have been obtained after 
$4\times4$ binning of the X-ray image. The position
of the centroid from \citet{1999ApJS..125..409C}
is indicated by the cross.}
\label{fig:4303con}
\end{figure}

\begin{table*} 
\scriptsize
\caption{Parameters from spectral fits.
Shown from left to right are (1) name,
(2) XSPEC model, (3) Galactic column density,
(4) intrinsic column density from XSPEC,
(5) photon index for power law
component, (6) peak energy of gaussian emission line, (7) equivalent width of the line, 
(8) plasma temperature, (9) \cs/degrees of freedom, (10) unabsorbed \lx\
from power law and (11) from thermal component (both in units ($10^{40}$\lunits).  The
spectral models are: PL=power law continuum; M=MEKAL thermal plasma
(solar abundances); RS=Raymond-Smith thermal plasma (solar
abundances); G=Gaussian emission line; W=photoelectric absorption
(Wisconsin cross-sections); ZW=redshifted W.  Uncertainties give 90\%
confidence intervals.}
\label{tab:spec} 
\centering 
\begin{tabular}{ll cc cc cc cc c } 
\hline\hline 
       Name    & Model                         & \nhgal\           & \nhint\                &    $\Gamma$         &      Fe K$\alpha$      &        EW              & $kT$                  &\cs/d.o.f.& \lx(PL)          & \lx(th)                          \\ 
               &                               &($10^{20}$ \cunits)&($10^{22}$ \cunits)     &                     &                        &       (keV)            & ($0.1$ keV)           &          &                  &                                  \\
\hline 

NGC 2782 & (PL+M+G)*ZW*W                  & 1.76             & $<0.07$                &\aer{0.0}{+0.6}{-0.3}&      6.4             &\aer{1.5}{+0.7}{-0.7}      &\aer{9.62}{+0.19}{-0.07}     &32.3/31   &\aer{4.81}{+3.17}{-2.09}& \aer{0.59}{+0.08}{-0.15}   \\




NGC 3310  &(PL+M+G)*ZW*W                & 1.13              &\aer{0.23}{+0.05}{-0.05}  & \aer{1.3}{+0.2}{-0.3}&     6.4                &\aer{0.3}{+0.2}{-0.2}  & \ldots\                     &68.6/83   &\aer{1.95}{+1.69}{-0.93}& \aer{0.42}{+0.70}{-0.35}    \\

NGC 4102 & (PL+RS+G)*ZW*W                 & 1.77             &\aer{0.19}{+0.33}{-0.65}&\aer{1.7}{+0.9}{-1.3}&    \ldots\             &        \ldots\          &\aer{0.7}{+0.2}{-0.4}        &18.6/11   &\aer{1.52}{+1.08}{-1.25}&\aer{2.99}{+11.94}{-2.78}     \\

NGC 4303  & (PL+M)*ZW*W                        & 1.67              & $<0.03$                &\aer{1.9}{+0.3}{-0.2}&       \ldots\          &    \ldots\           &\aer{5.37}{+0.09}{-0.06} &35.0/31   &\aer{0.17}{+0.06}{-0.03}& \aer{0.14}{+0.03}{-0.02}      \\


\hline 
\end{tabular} 
\end{table*}

\begin{table} 
\scriptsize
\caption{H1 and H2 results. Shown are \nhint-$\Gamma$ pairs which, assuming a
pure power-law spectrum, reproduce the \x\ colours calculated from the X-ray
image for sources for which no spectra
have been extracted. From left to right columns give (1) name, (2) Galactic column density, 
(3) intrinsic column density, and
(4) photon index for power law component. The last column (5) indicates
the type of object consistent with the \x\ colours, from 
a comparison with Fig. 4 of \citet{2003ApJ...595..719P} (see \fr{fig:prest}).
As mentioned in the text, there are two
alternatives for NGC 2146.}
\label{tab:hr} 
\centering 
\begin{tabular}{lr rr r} 
\hline\hline
Name                            & \nhgal\           &\nhint\            & $\Gamma$  & Type   \\
                                &($10^{20}$ \cunits)&($10^{22}$ \cunits)&           &        \\
\hline
NGC 278U                        &13.3              &0         & 2.7   & LMXB       \\
NGC 278L                        &13.3              &0         & 2.3   & LMXB       \\

NGC 2146                         &7.30              &4.0      & 1.8   & AGN        \\
                                 &                  &1.0      & 0.3   & HMXB       \\

NGC 4217U                        &1.23              &1.3      & 1.0   & HMXB       \\
NGC 4217L                        &1.23              &2.5      & 1.0   & HMXB       \\

NGC 5905                        & 1.44             & 0        & 2.5   & LMXB       \\


\hline
\end{tabular} 
\end{table}

\begin{figure}
\includegraphics[width=8cm,angle=0]{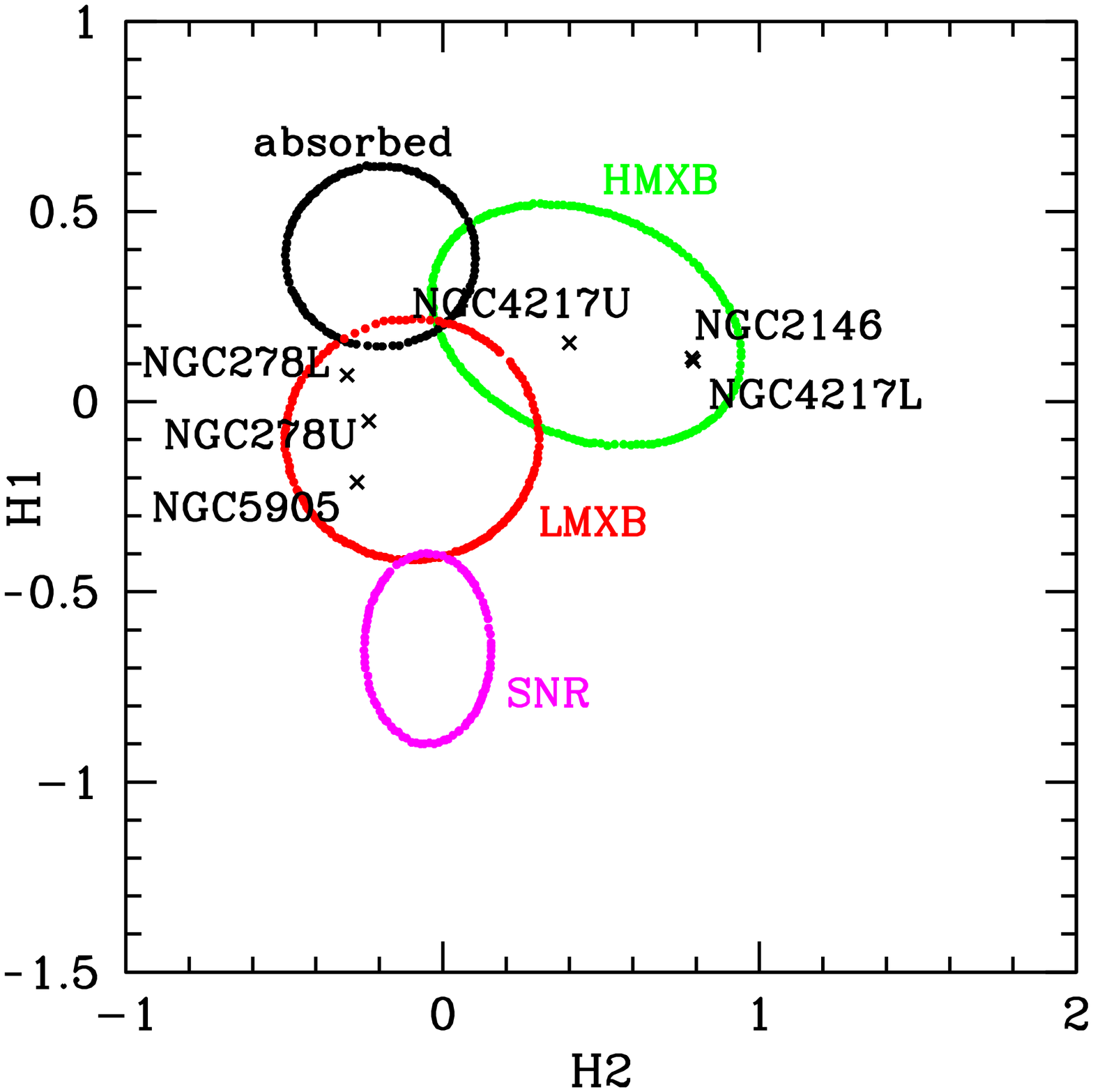}
\caption{\x\ colour-colour plot. The coloured contours,
taken from \citet{2003ApJ...595..719P}, indicate regions
for LMXBs (red), HMXBs (green), absorbed sources (black),
and thermal supernova remnants (magenta, SNR). The crosses
indicate calculated colours for sources in our sample for which we have
no spectra.}
\label{fig:prest}
\end{figure}

\end{document}